\documentclass[12pt]{iopart}
\usepackage{iopams}
\usepackage{verbatim}
\usepackage{amsmath}
\usepackage{amsfonts}
\usepackage{amssymb}
\usepackage{graphicx}
\usepackage{bm}
\usepackage{color}
\usepackage[dvipsnames]{xcolor}
\usepackage{hyperref}
\usepackage{ulem}
\usepackage{multirow}
\usepackage{braket}
\begin{document}

\title{Transitions of the Lyapunov spectrum and entanglement entropy in monitored quantum dynamics with homogeneous unitary gates}

\author{Ken Mochizuki$^{1,2}$ and Ryusuke Hamazaki$^{2,3}$}
\address{$^1$Department of Applied Physics, University of Tokyo, Tokyo 113-8656, Japan}
\address{$^2$Nonequilibrium Quantum Statistical Mechanics RIKEN Hakubi Research Team,
RIKEN Cluster for Pioneering Research (CPR), 2-1 Hirosawa, Wako 351-0198, Saitama, Japan}
\address{$^3$RIKEN Interdisciplinary Theoretical and Mathematical Sciences Program (iTHEMS), 2-1 Hirosawa, Wako 351-0198, Saitama, Japan}
\ead{ken-mochizuki@ap.t.u-tokyo.ac.jp}
\vspace{10pt}

\begin{abstract}
We explore the Lyapunov spectrum and entanglement entropy in quantum trajectories evolved by quantum measurements and spatially homogeneous unitary gates. 
In models with temporally random and Floquet unitary gates, we find that the Lyapunov exponents typically converge to values independent of measurement outcomes and that spectral transitions of the Lyapunov spectrum and entanglement transitions of ground states occur at the same measurement thresholds. 
Our results indicate that (i) randomness due to quantum measurements is enough, or equivalently, randomness of unitary gates is not essential, for the typical convergence of the Lyapunov spectrum and that (ii) the correspondence between the spectral transition and entanglement transition is ubiquitous in a wide range of systems. 
We also discuss why the typical convergence occurs, in light of the uniqueness and positive definiteness of stationary states in the completely-positive trace-preserving (CPTP) dynamics averaged over random measurement outcomes. 
Our discussions indicate that properties of the averaged CPTP dynamics play a key role in characterizing measurement-induced transitions in quantum trajectories. 
\end{abstract}

\section{Introduction}
\label{sec:introduction}
Understanding quantum phase transitions is one of the central themes in statistical physics. 
Different quantum phases are distinguished through various quantities, such as the entanglement entropy and correlation functions. 
In an isolated system, the spectral gap of the Hamiltonian describing the system is also a key ingredient in understanding the quantum phase transition. 
Indeed, the entanglement entropy of the ground state exhibits the logarithmic scaling in a gapless phase and the area-law scaling in a gapped phase, respectively, and the gap closing leads to quantum phase transitions \cite{hastings2006spectral,hastings2007area,zeng2019quantum}. 

Entanglement transitions in quantum systems exposed to quantum measurements have also been extensively explored in recent years. 
In such monitored quantum systems, the competition between unitary dynamics and quantum measurements, which respectively enhance and suppress entanglement growth, leads to measurement-induced entanglement transitions \cite{li2018quantum,
cao2019entanglement,
chan2019unitary,
li2019measurement,
skinner2019measurement,
szyniszewski2019entanglement,
bao2020theory,
choi2020quantum,
fuji2020measurement,
gullans2020dynamical,
lunt2020measurement,
szyniszewski2020universality,
tang2020measurement,
turkeshi2020measurement,
alberton2021entanglement,
lu2021spacetime,
agrawal2022entanglement,
barratt2022transitions,
block2022measurement,
minato2022fate,
muller2022measurement,
noel2022measurement,
sierant2022dissipative,
sierant2022universal,
zabalo2022operator,
granet2023volume,
koh2023measurement,
le2023volume,
loio2023purification,
majidy2023critical,
oshima2023charge,
yamamoto2023localization,
kumar2024boundary,
aziz2024critical,
mochizuki2025measurement}. 
Various aspects of measurement-induced entanglement transitions, such as invisibility in the averaged dynamics of density matrices \cite{li2018quantum,skinner2019measurement}, critical properties \cite{li2018quantum,
li2019measurement,
skinner2019measurement,
fuji2020measurement,
lunt2020measurement,
turkeshi2020measurement,
zabalo2022operator,
kumar2024boundary,
aziz2024critical}, and purification timescales \cite{gullans2020dynamical,
agrawal2022entanglement,
loio2023purification,
mochizuki2025measurement}, have been revealed. 
Moreover, the Lyapunov spectral analysis is emerging as an important tool to analyze measurement-induced transitions. 
Indeed, critical properties of spin systems \cite{zabalo2022operator,kumar2024boundary,aziz2024critical} and topological invariants in fermionic systems \cite{xiao2024topology,oshima2024topology} have been clarified through the Lyapunov analysis. 
There are two key features when we apply the Lyapunov analysis to quantum trajectories of weakly monitored many-body systems, where the probability distribution of measurement outcomes obeys the Born rule. 
One is the typical convergence of the Lyapunov exponents, which become independent of measurement outcomes almost surely in the long-time regime. 
This means that the Lyapunov exponents computed through one trajectory coincide with those averaged over many trajectories \cite{benoist2019invariant}. 
The other is that closing of the spectral gap of effective Hamiltonians introduced in the Lyapunov analysis leads to the entanglement transition of their ground states, which is analogous to quantum phase transitions in isolated systems. 
These two features have been observed in interacting spin systems where quantum states are evolved by spatiotemporally random unitary gates and measurements \cite{mochizuki2025measurement}.

In this work, we study the typical convergence and the transition of the many-body Lyapunov spectrum in quantum trajectories of monitored circuits with spatially homogeneous unitary gates, which are less random than the previously studied one \cite{mochizuki2025measurement}. 
We find that the Lyapunov exponenents show the typical convergence and that the spectral transitions occur upon the entanglement transitions in two models with or without temporal randomness in unitary gates.  
Our results indicate that randomness in the unitary gates is not essential, or equivalently, randomness of the quantum measurement is enough for the typical convergence and the relation between the spectrum and entanglement. 
We also discuss why the typical convergence occurs in our models, on the basis of stationary states in the completely-positive trace-preserving (CPTP) dynamics averaged over measurement outcomes. 
Our discussions indicate that properties of the averaged CPTP dynamics are important to characterize measurement-induced transitions in pure-state quantum trajectories, while the transitions are usually invisible in the averaged dynamics of density matrices. 

The rest of this paper is organized as follows. 
In Sec. \ref{sec:Lyapunov-analysis}, we review the Lyapunov spectral analysis. 
In Sec. \ref{sec:numerics}, we show the typical convergence of the Lyapunov spectrum, spectral transitions, and entanglement transitions in concrete models with spatially homogeneous unitary gates. 
In Sec. \ref{sec:typical-convergence}, we discuss the origin of the typical convergence of the Lyapunov spectrum.  
Section \ref{sec:summary} summarizes the manuscript.

\section{Lyapunov analysis}
\label{sec:Lyapunov-analysis}
We consider an $N \times N$ nonunitary random matrix $G(\omega)$ determined through a random variable $\omega$. 
In monitored quantum systems explored in this work, $\omega$  corresponds to a set of measurement outcomes combined with random parameters of unitary gates.
Then, $G(\omega)$ describes the corresponding time-evolution operator composed of nonunitary Kraus operators for the measurements and unitary operators for quantum gates, whose concrete forms are given in Sec. \ref{sec:numerics}. 
We consider products of random matrices,
\begin{align}
    V({\bm \omega}_t)=G(\omega_t)G(\omega_{t-1}) \cdots G(\omega_1)
    \label{eq:time-evolution_operator}
\end{align} 
with ${\bm \omega}_t=(\omega_1,\omega_2,\cdots,\omega_t)$,  where $\omega_t$ is a random variable at the time step $t$. 

We focus on the Lyapunov spectrum $\{\varepsilon_i({\bm \omega}_t)\}$ defined through the eigenequation
\begin{align}
    K({\bm \omega}_t)\ket{\Psi_i({\bm \omega}_t)}
    =\varepsilon_i({\bm \omega}_t)\ket{\Psi_i({\bm \omega}_t)}
    \label{eq:eigen-equation}
\end{align}
with $i=1,2,\cdots,N$, where the effective Hamiltonian $K({\bm \omega}_t)$ is
\begin{align}
    K({\bm \omega}_t)=-\frac{1}{2t}
    \ln\left[V({\bm \omega}_t)
    V^\dagger({\bm \omega}_t)\right].
\end{align} 
In other words, the Lyapunov spectrum is obtained through the singular-value decomposition of $V({\bm \omega}_t)$,
\begin{align}
    V({\bm \omega}_t)=\sum_i\Lambda_i({\bm \omega}_t)
    \ket{\Psi_i({\bm \omega}_t)}\bra{\Phi_i({\bm \omega}_t)},\ \ 
    \Lambda_i({\bm \omega}_t)
    =\exp\left[-\varepsilon_i({\bm \omega}_t)t\right],
    \label{eq:singular-value_decomposition}
\end{align}
where $\ket{\Psi_i({\bm \omega}_t)}$ and $\ket{\Phi_i({\bm \omega}_t)}$ are eigenvectors of $V({\bm \omega}_t)V^\dagger({\bm \omega}_t)$ and $V^\dagger({\bm \omega}_t)V({\bm \omega}_t)$, respectively:
\begin{align}
    V({\bm \omega}_t)V^\dagger({\bm \omega}_t)\ket{\Psi_i({\bm \omega}_t)}
    &=\left[\Lambda_i({\bm \omega}_t)\right]^2\ket{\Psi_i({\bm \omega}_t)},\\
    V^\dagger({\bm \omega}_t)V({\bm \omega}_t)\ket{\Phi_i({\bm \omega}_t)}
    &=\left[\Lambda_i({\bm \omega}_t)\right]^2\ket{\Phi_i({\bm \omega}_t)}.
\end{align} 
Here, they satisfy the orthonormality condition $\langle\Psi_i({\bm \omega}_t)|\Psi_j({\bm \omega}_t)\rangle=\langle\Phi_i({\bm \omega}_t)|\Phi_j({\bm \omega}_t)\rangle=\delta_{ij}$. 
In the following, we assume that there is no degeneracy of $\{\varepsilon_i({\bm \omega}_t)\}$ and array the Lyapunov exponents and the singular values such that  $\varepsilon_i({\bm \omega}_t)<\varepsilon_{i+1}({\bm \omega}_t)$ and $\Lambda_i({\bm \omega}_t)>\Lambda_{i+1}({\bm \omega}_t)$ are satisfied. 
In the long-time regime, the Lyapunov exponents exhibit the typical convergence, i.e. they become independent of outcomes ${\bm \omega}_t$,
\begin{align}
    \lim_{t\rightarrow\infty}\varepsilon_i({\bm \omega}_t)
    =\lim_{t\rightarrow\infty}-\frac{1}{t}\ln\left[\Lambda_i({\bm \omega}_t)\right]
    =\varepsilon_i
    \label{eq:Lyapunov-spectrum_long-time}
\end{align}
almost surely, when the monitored quantum system satisfies several conditions \cite{benoist2019invariant}, which is discussed in Sec. \ref{sec:typical-convergence}.

When we numerically compute the Lyapunov spectrum, it is difficult to obtain $\{\Lambda_i({\bm \omega}_t)\}$ and $\{\ket{\Psi_i({\bm \omega}_t)}\}$ through diagonalizing $V({\bm \omega}_t)V^\dagger({\bm \omega}_t)$ directly. 
This is because the singular values typically exhibit exponential decay as $t$ becomes large, which we can understand from Eqs. (\ref{eq:singular-value_decomposition}) and (\ref{eq:Lyapunov-spectrum_long-time}), and thus the values of $\Lambda_i({\bm \omega}_t)$ easily deviate from the numerical precision. 
We note that $0\leq\varepsilon_i({\bm \omega}_t)$ is always satisfied in monitored quantum systems. 
In addition, diagonalizing the $N \times N$ matrix $V({\bm \omega}_t)V^\dagger({\bm \omega}_t)$ requires huge numerical cost when the matrix size $N$ is large. 

On the other hand, we can efficiently compute $\{\varepsilon_i({\bm \omega}_t)\}$ and $\{\ket{\Psi_i({\bm \omega}_t)}\}$ with $i=1,2,\cdots,q \ll N$ using the Gram-Schmidt orthogonalization \cite{geist1990comparison,ershov1998concept}, which is explained below. 
First, we prepare $q$ different initial states $\ket{\tilde{\Psi}_i^0}$ that are orthonormalized as $\langle\tilde{\Psi}_i^0|\tilde{\Psi}_j^0\rangle=\delta_{ij}$. 
Through long-time dynamics by $V({\bm \omega}_t)$ and the Gram-Schmidt orthonormalization,  $\ket{\tilde{\Psi}_i({\bm \omega}_t)}$ approaches $\ket{\Psi_i({\bm \omega}_t)}$. 
To this end, we compute $b$ step dynamics of $\ket{\tilde{\Psi}_i({\bm \omega}_{sb})}$ generated by random matrices. 
We first consider
\begin{align}
    \ket{\varphi_i({\bm \omega}_{sb})}
    =V(\tilde{\bm \omega}_s)
    \ket{\tilde{\Psi}_i({\bm \omega}_{(s-1)b})},
    \label{eq:evolution_tilde}
\end{align}
where $\tilde{\bm \omega}_s=(\omega_{(s-1)b+1},\cdots,\omega_{sb})$, $V(\tilde{\bm \omega}_s)=G(\omega_{sb})G(\omega_{sb-1}) \cdots G(\omega_{(s-1)b+1})$, $s=1,2,\cdots$, and $\ket{\tilde{\Psi}_i({\bm \omega}_0)}=\ket{\tilde{\Psi}_i^0}$. 
If $b$ is not so large, singular values of $V(\tilde{\bm \omega}_s)$ can be within the numerical precision, and thus we can avoid the numerical breakdown. 
At each $s$, we carry out the Gram-Schmidt orthogonalization of $\ket{\varphi_i({\bm \omega}_{sb})}$,
\begin{align}
    \ket{\chi_i({\bm \omega}_{sb})}
    =\left[I-\Pi_i({\bm \omega}_{sb})\right]
    \ket{\varphi_i({\bm \omega}_{sb})},
    \label{eq:orthogonalization}
\end{align}
where $I$ is the identity operator. 
Here, $\Pi_i({\bm \omega}_{sb})=\sum_{j=1}^{i-1}\ket{\tilde{\Psi}_j({\bm \omega}_{sb})}\bra{\tilde{\Psi}_j({\bm \omega}_{sb})}$ with $i\geq2$ is the projection operator onto the Hilbert space spanned by $\{\ket{\tilde{\Psi}_j({\bm \omega}_{sb})}\}_{j=1}^{i-1}$, and $\Pi_i({\bm \omega}_{sb})=0$ for $i=1$. 
Then, the candidate of $\ket{\Psi_i({\bm \omega}_t)}$ at $t=sb$ becomes
\begin{align}
    \ket{\tilde{\Psi}_i({\bm \omega}_{sb})}
    =\frac{\ket{\chi_i({\bm \omega}_{sb})}}
    {\sqrt{\langle\chi_i({\bm \omega}_{sb})
    |\chi_i({\bm \omega}_{sb})\rangle}}.
  \label{eq:normalization}
\end{align}
In the procedure, we obtain $\ket{\tilde{\Psi}_i({\bm \omega}_{sb})}$ from $\{\ket{\tilde{\Psi}_j({\bm \omega}_{sb})}\}$ with $j=1,2,\cdots,i-1$ and this is iteratively carried out from $i=1$ to $i=q$.
The candidate of the $i$th Lyapunov exponent becomes 
\begin{align}
    \tilde{\varepsilon}_i({\bm \omega}_{sb})=-\frac{1}{sb}\sum_{r=1}^s
    \ln\left[\sqrt{\langle\chi_i({\bm \omega}_{rb})|\chi_i({\bm \omega}_{rb})\rangle}\right].
    \label{eq:Lyapunov-spectrum_each_separate}
\end{align}
It is known that $\tilde{\varepsilon}_i({\bm \omega}_{sb})$ and $\ket{\tilde{\Psi}_i({\bm \omega}_{sb})}$ respectively approach $\varepsilon_i({\bm \omega}_{sb})$ and $\ket{\Psi_i({\bm \omega}_{sb})}$,
\begin{align}
    \tilde{\varepsilon}_i({\bm \omega}_{sb})
    &\rightarrow\varepsilon_i({\bm \omega}_{sb}),
    \label{eq:spectrum-candidates_limit}\\
    \ket{\tilde{\Psi}_i({\bm \omega}_{sb})}
    &\rightarrow\ket{\Psi_i({\bm \omega}_{sb})},
    \label{eq:state-candidates_limit}
\end{align}
for sufficiently large $s$ \cite{ershov1998concept}. 
Justifications of the abovementioned procedure based on the Gram-Schmidt orthonormalization are reviewed in \ref{sec:justification_spectrum} and \ref{sec:justification_state}.

\section{Results in concrete models}
\label{sec:numerics}
\subsection{Monitored quantum dynamics}
\label{sucsec:mnitored-curcuits}
\begin{figure}[tb]
\begin{center}
    \includegraphics[width=15cm]{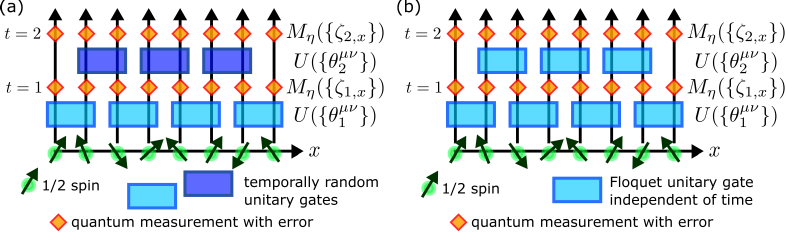}
\caption{Schematic figures of the monitored quantum dynamics where spins $1/2$ are evolved by unitary gates and quantum measurements described by $U(\{\theta^{\mu\nu}_t\})$ and $M_\eta(\{\zeta_{t,\ell }\})$, respectively. 
The local unitary gates are spatially homogeneous. 
(a) In the temporally random model, the parameters $\{\theta_t^{\mu\nu}\}$ describing the unitary dynamics are randomly chosen at each time step $t$. 
(b) In the Floquet model, the parameters are independent of $t$ and thus the unitary dynamics is temporally homogeneous.}
\label{fig:circuit}
\end{center}
\end{figure}
We consider monitored quantum spin-$1/2$ systems where spins are arrayed on a one-dimensional chain, as depicted in Fig. \ref{fig:circuit}. 
Quantum states are evolved by unitary gates and generalized measurements on spins. 
Our unitary dynamics are composed of local unitary gates acting on spins at positions $\ell $ and $\ell +1$, 
\begin{align}
    \mathsf{U}_{\ell ,\ell +1}(\{\theta_t^{\mu\nu}\})
    =\exp\left[-iH_{\ell ,\ell +1}(\{\theta_t^{\mu\nu}\})\right],\ \ 
    H_{\ell ,\ell +1}(\{\theta_t^{\mu\nu}\})=\sum_{\mu,\nu}\theta_t^{\mu\nu}\sigma_\ell ^\mu\sigma_{\ell +1}^\nu
    \label{eq:unitary-operator_each}
\end{align}
where $\mu,\nu=0,1,2,3$ are indices specifying Pauli matrices, $\sigma_\ell ^\mu$ is the $\mu$th Pauli matrix at a position $\ell \in[1,L]$, $L$ is the number of spins, and $\{\theta_t^{\mu\nu}\}$ are parameters that determine unitary dynamics. 
We impose the open boundary condition $H_{L,L+1}(\{\theta_t^{\mu\nu}\})=0$. 
We focus on spatially homogeneous unitary gates, where $\{\theta_t^{\mu\nu}\}$ are independent of $\ell $. 
The unitary dynamics at time step $t$ becomes
\begin{align}
    U(\{\theta_t^{\mu\nu}\})=\begin{cases}
    \prod_{\ell :\mathrm{odd}}\mathsf{U}_{\ell ,\ell +1}(\{\theta_t^{\mu\nu}\})&
    (t:\mathrm{odd})\\
    \prod_{\ell :\mathrm{even}}\mathsf{U}_{\ell ,\ell +1}(\{\theta_t^{\mu\nu}\})&
    (t:\mathrm{even})
    \end{cases}.
    \label{eq:unitary-operator_total}
\end{align}
There are two models with different parameters for unitary dynamics, (i) the temporally random model where $\{\theta_t^{\mu\nu}\}$ are randomly distributed at each step $t$ and (ii) the Floquet model where $\{\theta_t^{\mu\nu}\}$ are constant values independent of $t$. They are schematically illustrated in Fig. \ref{fig:circuit} (a) and (b).  

After the unitary dynamics by $U(\{\theta_t^{\mu\nu}\})$, generalized quantum measurements are carried out at all local spins. 
Our Kraus operators describing the measurement of the spin at a site $\ell $ are
\begin{align}
    \mathsf{M}_\eta(\zeta_{t,\ell }=\pm)
    =\frac{1}{2}\left[\left(\sqrt{\frac{1}{2}+\eta}+\sqrt{\frac{1}{2}-\eta}\right)\sigma_\ell ^0 \pm\left(\sqrt{\frac{1}{2}+\eta}-\sqrt{\frac{1}{2}-\eta}\right)\sigma_\ell ^3\right],
    \label{eq:Kraus-operator_each}
\end{align}
where we write the measurement outcome at time $t$ and position $\ell $ as $\zeta_{t,\ell }=\pm$.
The Kraus operators satisfy the trace-preserving condition for conservation of probability, $\sum_{\zeta_{t,\ell }=\pm}\mathsf{M}_\eta^\dagger(\zeta_{t,\ell })\mathsf{M}_\eta(\zeta_{t,\ell })=\sigma_\ell ^0$, which is the identity operator. 
The parameter $\eta\in[0,1/2]$ represents the strength of measurement; $\eta=0$ corresponds to no measurement and $\eta=1/2$ corresponds to the projective measurement. 
Since the generalized measurements are carried out at all sites, we consider the Kraus operators for the entire system at each time,
\begin{align}
    M_\eta(\{\zeta_{t,\ell }\})=\prod_{\ell =1}^L
        \mathsf{M}_\eta(\zeta_{t,\ell }),
    \label{eq:Kraus-operator_total}
\end{align}
where $\{\zeta_{t,\ell }\}=(\zeta_{t,1},\zeta_{t,2},\cdots,\zeta_{t,L})$ is the set of measurement outcomes for all sites at time step $t$. 
In the following, to simplify the notation, we describe dependence on $\{\zeta_{t,\ell }\}$ and $\{\theta_t^{\mu\nu}\}$ through the combined variable $\omega_t=(\{\zeta_{t,\ell }\},\{\theta_t^{\mu\nu}\})$. 
When the sequence of random variables from the first to $t$ steps becomes ${\bm \omega}_t=(\omega_1,\omega_2,\cdots,\omega_t)$, a quantum pure state is evolved as
\begin{align}
    \ket{\psi_\eta({\bm \omega}_t)}\propto 
    V_\eta({\bm \omega}_t)\ket{\psi_0},\ \ 
    V_\eta({\bm \omega}_t)=M_\eta(\{\zeta_{t,\ell }\})U(\{\theta_t^{\mu\nu}\})
     \cdots M_\eta(\{\zeta_{1,\ell }\})U(\{\theta_1^{\mu\nu}\}),
    \label{eq:dynamics}
\end{align}
where $\ket{\psi_0}$ is a randomly chosen initial state. 
The nonunitary time-evolution operator in Sec. \ref{sec:Lyapunov-analysis} corresponds to $G(\omega_t)=M_\eta(\{\zeta_{t,\ell }\})U(\{\theta_t^{\mu\nu}\})$. 
When the quantum state at $t$ is $\ket{\psi_\eta({\bm \omega}_t)}$, the probability that measurement outcomes at $t+1$ become $\{\zeta_{t+1,\ell }\}$ is given by the Born rule,
\begin{align}
    P_\eta(\{\zeta_{t+1,\ell }\}|\{\theta_{t+1}^{\mu\nu}\},{\bm \omega}_t)
    =\left|M_\eta(\{\zeta_{t+1,\ell }\})U(\{\theta_{t+1}^{\mu\nu}\})
    \ket{\psi_\eta({\bm \omega}_t)}\right|^2,
    \label{eq:probability}
\end{align}
under the condition that parameters for unitary dynamics at the step $t+1$ are $\{\theta_{t+1}^{\mu\nu}\}$. 

\subsection{Quantities that we focus on}
\label{sucsec:spectrum_entanglement}
In the above nonunitary random dynamics, we explore the Lyapunov spectrum explained in Sec. \ref{sec:Lyapunov-analysis}. 
In our models explored below, we numerically find that the Lyapunov exponents exhibit the typical convergence, i.e. they become independent of measurement outcomes almost surely,
\begin{align}
    \varepsilon_{i,\eta}^L
    =\lim_{t\rightarrow\infty}\varepsilon_{i,\eta}^L({\bm \omega}_t).
    \label{eq:Lyapunov-spectrum_long-time_eta}
\end{align}
Here, we attach the subscript $\eta$ and superscript $L$ since we explore how the Lyapunov exponents depend on the measurement strength $\eta$ and system size $L$. 
To check the convergence of $\tilde{\varepsilon}_{i,\eta}^L({\bm \omega}_t)$ to $\varepsilon_{i,\eta}^L$, we take the time average of $\tilde{\varepsilon}_{i,\eta}^L({\bm \omega}_{sb})$ over $f$ points,
\begin{align}
    \mathbb{E}_t\left[\tilde{\varepsilon}_{i,\eta}^L({\bm \omega}_{sb})\right]
    =\frac{1}{f}\sum_{r=0}^{f-1}
    \tilde{\varepsilon}_{i,\eta}^L({\bm \omega}_{(s-r)b}).
    \label{eq:time-average}
\end{align}
In actual calculations, we choose $f$ larger than $200$ to ensure the numerical accuracy. 
If the step $s$ satisfies $s\geq2f$ and the variance $\mathbb{V}_t\left[\tilde{\varepsilon}_{i,\eta}^L({\bm \omega}_{sb})\right]=\mathbb{E}_t\left[\tilde{\varepsilon}_{i,\eta}^L({\bm \omega}_{sb})^2\right]-\mathbb{E}_t\left[\tilde{\varepsilon}_{i,\eta}^L({\bm \omega}_{sb})\right]^2$ satisfies $\sqrt{\mathbb{V}_t\left[\tilde{\varepsilon}_{i,\eta}^L({\bm \omega}_{sb})\right]}/\mathbb{E}_t\left[\tilde{\varepsilon}_{i,\eta}^L({\bm \omega}_{sb})\right]\leq d$, where $d$ is a threshold smaller than $10^{-2}$, we regard the average $\mathbb{E}_t\left[\tilde{\varepsilon}_{i,\eta}^L({\bm \omega}_{sb})\right]$ as the $i$th Lyapunov exponent $\varepsilon_{i,\eta}^L$.  
Note that we empirically find that the time average $\mathbb{E}_t\left[\varepsilon_{i,\eta}^L({\bm \omega}_{sb})\right]$ in Eq. (\ref{eq:time-average}) depends on $b$ in the procedure explained in Sec. \ref{sec:Lyapunov-analysis} if we choose small $b$ such as $b=2,4,8$, which is considered to be a numerical artifact. 
This dependence on $b$ becomes stronger as $\eta$ becomes smaller. 
Thus, at several $\eta$, we check results with different $b$ and determine $b$ large enough such that results become independent of $b$. 
For example, in the temporally random model with $\eta=0.1$ and $L=18$, we have confirmed that the Lyapunov spectrum becomes almost independent of $b$ in the range $32\leq b \leq 256$. We note that, in this range, the numerical breakdown explained in Sec. \ref{sec:Lyapunov-analysis} does not occur. 

As an indicator of measurement-induced transitions, we focus on the spectral gap given by
\begin{align}
    \Delta_\eta^L
    =\varepsilon_{2,\eta}^L-\varepsilon_{1,\eta}^L.
    \label{eq:gap}
\end{align}
The spectral gap in the thermodynamic limit,
\begin{align}
    \Delta_\eta=\lim_{L\rightarrow\infty}\Delta_\eta^L,
\end{align}
exhibits a transition when we vary the measurement strength, which is first discussed in Ref.~\cite{mochizuki2025measurement}. 
To compute $\Delta_\eta$, we extrapolate the numerical data $\{\Delta_\eta^L\}$ to the thermodynamic limit $L\rightarrow\infty$ employing the fitting function
\begin{align}
    \tilde{\Delta}^L(\alpha_\eta,\beta_\eta,\gamma_\eta)=\gamma_\eta
    +\exp(\alpha_\eta-L/\beta_\eta).
\end{align}
For various values of $\eta$, we sweep $\gamma_\eta$ from $-\min_L\Delta_\eta^L$ to $+\min_L\Delta_\eta^L$. 
At each $\gamma_\eta$, we perform the linear least-square fitting based on the cost function $\delta(\alpha_\eta,\beta_\eta,\gamma_\eta)=\sum_L\left(\ln\left[\frac{\tilde{\Delta}^L(\alpha_\eta,\beta_\eta,\gamma_\eta)-\gamma_\eta}{\Delta_\eta^L-\gamma_\eta}\right]\right)^2=\sum_L\left[\ln\left(\Delta_\eta^L-\gamma_\eta\right)-(\alpha_\eta-L/\beta_\eta)\right]^2$. 
Then, we have $\delta_\mathrm{m}(\gamma_\eta)$ as a function of $\gamma_\eta$, which is the minimum value of $\delta(\alpha_\eta,\beta_\eta,\gamma_\eta)$ obtained through finding best $\alpha_\eta$ and $\beta_\eta$. 
Among the obtained values, $\gamma_\eta$ such that minimizes $\delta_\mathrm{m}(\gamma_\eta)$ is regarded as $\Delta_\eta$. 

We also explore the entanglement entropy of the ground states of the effective Hamiltonians $K_\eta({\bm\omega}_t)$,
\begin{align}
    S^A\left[\Psi_{1,\eta}({\bm \omega}_t)\right]=-\mathrm{tr}\left[\rho^A\left[\Psi_{1,\eta}({\bm \omega}_t)\right]\ln\left(\rho^A\left[\Psi_{1,\eta}({\bm \omega}_t)\right]\right)\right],
    \label{eq:entanglement-entropy}
\end{align}
where the reduced density matrix of a state $\ket{\psi}$ for a subsystem $A$ is given by
\begin{align}
    \rho^A(\psi)=\mathrm{tr}_{\bar{A}}\left(
    \ket{\psi}\bra{\psi}\right)
    \label{eq:reduced-density-matrix}
\end{align}
with $\bar{A}$ being the complement of the subsystem $A$. 
It is known that the threshold of the entanglement transition can be detected through a peak of the mutual information, 
\begin{align}
    I^{A,B}\left[\Psi_{1,\eta}({\bm \omega}_t)\right]
    =S^A\left[\Psi_{1,\eta}({\bm \omega}_t)\right]
    +S^B\left[\Psi_{1,\eta}({\bm \omega}_t)\right]
    -S^{AB}\left[\Psi_{1,\eta}({\bm \omega}_t)\right].
    \label{eq:mutual-information}
\end{align}
This is because the mutual information gives the upper bound of correlation functions of observables in regions $A$ and $B$ \cite{li2019measurement,wolf2008area}. 

Reference~\cite{mochizuki2025measurement} pointed out that the spectral transition and the entanglement transition coincide in monitored quantum circuits, which reminds us of the ground-state phase transitions in isolated quantum systems \cite{zeng2019quantum}. 
However, such a coincidence has been confirmed in monitored systems with spatially and temporally random unitary gates, and it is important to confirm the coincidence for our monitored circuits without such randomness.

\subsection{Numerical results}
We explore the Lyapunov spectrum, entanglement entropy, and mutual information in the two models. 
In both models, we find that the Lyapunov spectrum exhibits the typical convergence, that the spectral gap and ground-state entanglement entropy exhibit transitions, and that their thresholds correspond. 

\label{sucsec:numerical results}
\begin{figure}[tb]
\begin{center}
    \includegraphics[width=14cm]{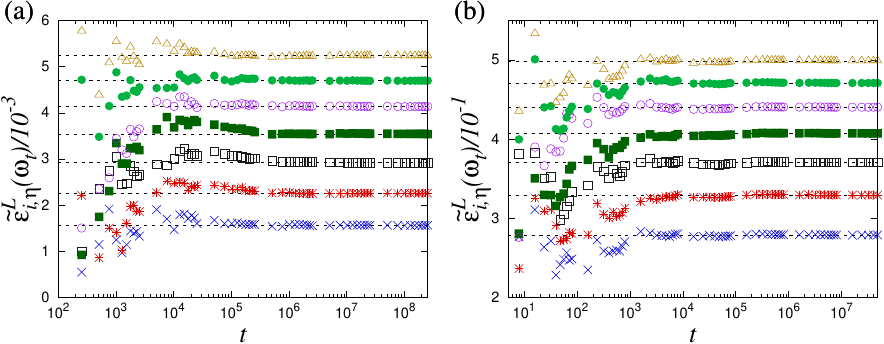}
\caption{The Lyapunov exponents with $i=2,\cdots,8$ in the temporally random model, evaluated through the difference from $\tilde{\varepsilon}_{1,\eta}^L({\bm \omega}_t)$, where the strengths of the measurement are (a) $\eta=0.11$ and (b) $\eta=0.36$. 
The system size is $L=12$ in both figures, and the time bins are (a) $b=256$ and (b) $b=8$. 
The black dashed lines represent sample averages of $\tilde{\varepsilon}_{i,\eta}^L({\bm \omega}_t)-\tilde{\varepsilon}_{1,\eta}^L({\bm \omega}_t)$ over $100$ trajectories, at (a) $t=2304000$ and (b) $t=80000$. }
\label{fig:spectrum_temporally-random}
\end{center}
\end{figure}

The first model is the temporally random and spatially homogeneous model, where $\{\theta_t^{\mu\nu}\}$ randomly depend on $t$ but not on $\ell $. 
The schematic picture is shown in Fig. \ref{fig:circuit} (a). 
At each step $t$, the random values of $\{\theta_t^{\mu\nu}\}$ are chosen from the box distribution whose range is
\begin{align}
    \theta^{\mu\nu}_t\in[-\pi,+\pi].
    \label{eq:box-distribution}
\end{align}
Figure \ref{fig:spectrum_temporally-random} shows the Lyapunov exponents up to $i=8$ with (a) $\eta=0.11$ and (b) $\eta=0.36$. 
All Lyapunov exponents computed through the procedure explained in Sec. \ref{sec:Lyapunov-analysis} converge to constant values independent of $t$. 
In both Fig. \ref{fig:spectrum_temporally-random} (a) and (b), the convergence values obtained from one trajectory coincide with those averaged over various trajectories, which indicates that Eq. (\ref{eq:Lyapunov-spectrum_long-time_eta}) is satisfied. 
We note that it takes longer time for the Lyapunov exponents to converge to constants, and thus the computational cost becomes heavier, as the measurement strength $\eta$ becomes smaller and the systems size $L$ becomes larger.

\begin{figure}[tb]
\begin{center}
    \includegraphics[width=14cm]{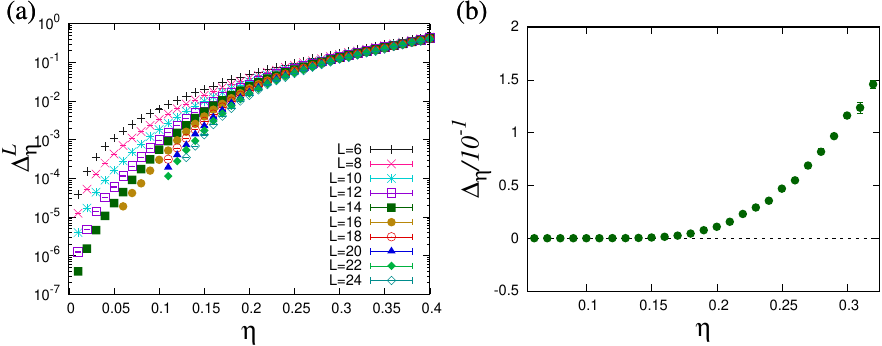}
\caption{Spectral gaps in the temporally random model. 
(a) Numerically obtained $\Delta_\eta^L$ as functions of the measurement strength $\eta$ for various $L$. 
(b) The spectral gap in the thermodynamic limit, $\Delta_\eta$, obtained through the extrapolation of the numerical data. }
\label{fig:gap_temporally-random}
\end{center}
\end{figure}

Figure \ref{fig:gap_temporally-random} (a) shows the spectral gap $\Delta_\eta^L$ for various system sizes $L$ as functions of the measurement strength $\eta$. 
We find that $\Delta_\eta^L$ becomes almost independent of $L$ in the large $\eta$ regime, which indicates that the gapped phase is realized when measurements are akin to projective measurements. 
On the other hand, in the small $\eta$ regime where measurements are weak, the gaps are decreasing functions of $L$ and thus the gapless phase is realized. 
We can also confirm the spectral transition from $\Delta_\eta$ as shown in Fig. \ref{fig:gap_temporally-random} (b).
We find that $\Delta_\eta$ is near $0$ and almost flat for $\eta\lesssim0.18$, whereas it becomes an increasing function of $\eta$ for $0.2\lesssim\eta$. 

\begin{figure}[tb]
\begin{center}
    \includegraphics[width=15cm]{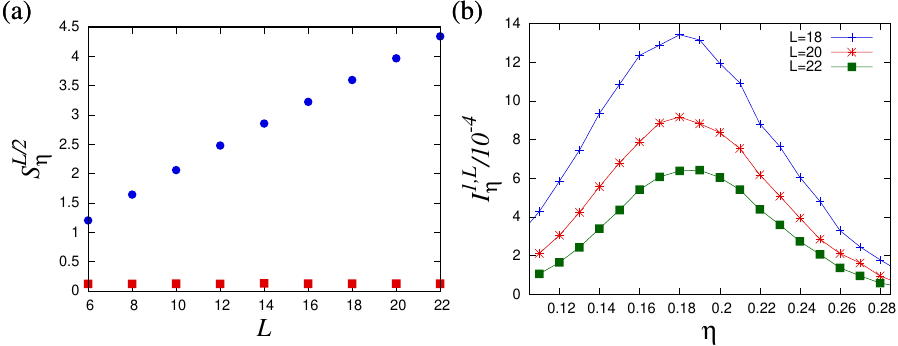}
\caption{The entanglement entropy and mutual information for the temporally random model. 
(a) The half-chain entanglement entropy $S^{L/2}_\eta$ as functions of $L$ with $\eta=0.11$ (blue circles) and $\eta=0.36$ (red squares). 
The time average is taken over $10^4$ steps after $t$ becomes larger than $\left|\ln(10^{-3})/\Delta_\eta^L\right|$. 
(b) The mutual information as functions of $\eta$, where the time average is taken over $10^5$ steps after $\left|\ln(10^{-3})/\Delta_\eta^L\right|$. }
\label{fig:entanglement_temporally-random}
\end{center}
\end{figure}

We next compare the spectral transition and the ground-state entanglement transition of the effective Hamiltonian. 
Figure \ref{fig:entanglement_temporally-random} (a) shows system-size dependence of the half-chain entanglement entropy $S^{L/2}_\eta$, which is the average of $S^{L/2}\left[\Psi_{1,\eta}({\bm \omega}_t)\right]$ over time.  
In the gapped phase with large $\eta$, $S^{L/2}_\eta$ becomes almost independent of $L$ as shown in Fig. \ref{fig:entanglement_temporally-random} (a), which indicates that the gapped phase corresponds to the area-law entanglement phase of ground states. 
On the other hand, in the gapless phase with small $\eta$, $S^{L/2}_\eta$ is proportional to $L$ as shown in Fig. \ref{fig:entanglement_temporally-random} (a), indicating the volume-law entanglement phase for $\ket{\Psi_{1,\eta}({\bm \omega}_t)}$. 
As shown in Fig. \ref{fig:entanglement_temporally-random} (b), a peak of the mutual information $I^{1,L}_\eta$, which is the time average of $I^{1,L}\left[\Psi_{1,\eta}({\bm \omega}_t)\right]$, exists at $\eta_\mathrm{c}$ in the range $0.18\leq\eta_\mathrm{c}\leq0.2$, while there is a slight shift of the peak as $L$ is changed due to finite-size effects. 
The peak of the mutual information, which corresponds to the entanglement transition, is consistent with the spectral transition from the gapped phase to the gapless phase, as shown in Fig. \ref{fig:gap_temporally-random} (b). 

The second model we study is the monitored system with the Floquet unitary gates, where $\{\theta^{\mu\nu}_t\}$ are independent of $t$; we simply call this model as the Floquet model, although the dynamics includes temporal randomness due to measurements. 
The schematic picture is shown in Fig. \ref{fig:circuit} (b). 
We focus on a parameter set
\begin{align}
    \theta^{00}=0,\ \theta^{11}=0.71\pi,\ &\theta^{22}=1.43\pi,\ \theta^{33}=0.27\pi,
    \label{eq:theta_1}\\ 
    \theta^{10}=1.21\pi,\ \theta^{01}=0.43\pi,\ 
    \theta^{20}=0.83\pi,\ &\theta^{02}=0.62\pi,\ 
    \theta^{30}=1.53\pi,\ \theta^{03}=0.47\pi,
    \label{eq:theta_2}\\ 
    \theta^{12}=0.35\pi,\ \theta^{21}=0.69\pi,\ 
    \theta^{23}=1.19\pi,\ &\theta^{32}=0.75\pi,\ 
    \theta^{31}=0.12\pi,\ \theta^{13}=1.87\pi,
    \label{eq:theta_3}
\end{align}
such that coefficients for all possible Pauli strings in $H_{\ell ,\ell +1}(\{\theta^{\mu\nu}\})$ are non-zero and these have no specific structure, which leads to the absence of strong symmetry. 
We note that the coefficient $\theta^{00}$ corresponding to the identity $\sigma_\ell^0\sigma_{\ell+1}^0$ is ignored since it has no effect on dynamics. 
As discussed in Sec. \ref{sec:typical-convergence}, the absence of strong symmetry is important when we characterize measurement-induced transitions through the Lyapunov spectrum. 
\begin{figure}[tb]
\begin{center}
    \includegraphics[width=14cm]{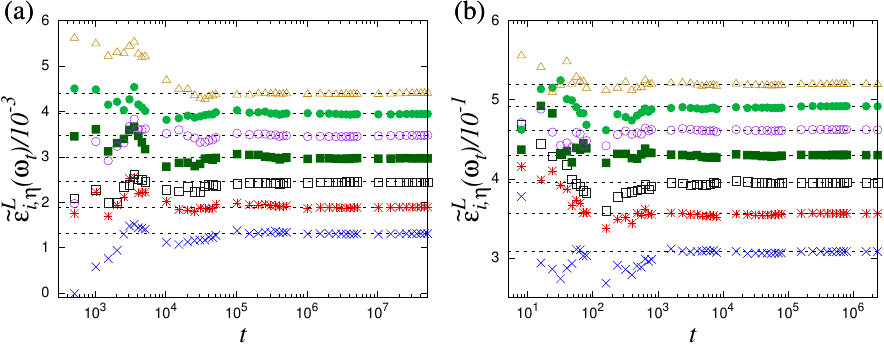}
\caption{The Lyapunov exponents with $i=2,\cdots,8$ for the monitored dynamics with the Floquet unitary gates, evaluated through the difference from $\tilde{\varepsilon}_{1,\eta}^L({\bm \omega}_t)$, where the measurement strengths are (a) $\eta=0.12$ and (b) $\eta=0.37$. 
The system size is $L=14$ in both figures, and the time bins are (a) $b=512$ and (b) $b=8$. 
The black dashed lines represent sample averages of $\tilde{\varepsilon}_{i,\eta}^L({\bm \omega}_t)-\tilde{\varepsilon}_{1,\eta}^L({\bm \omega}_t)$ over $100$ trajectories, at (a) $t=512000$ and (b) $t=16000$. }
\label{fig:spectrum_Floquet-model}
\end{center}
\end{figure}
Figure \ref{fig:spectrum_Floquet-model} shows the Lyapunov exponents up to $i=8$ with (a) $\eta=0.12$ and (b) $\eta=0.37$. 
In the same way as the temporally random model, all Lyapunov exponents obtained from one trajectory in the Floquet model converge to those averaged over many trajectories. 
This indicates that the spatial and temporal randomness of unitary gates have negligible effect on the typical convergence of the Lyapunov spectrum. 

\begin{figure}[tb]
\begin{center}
    \includegraphics[width=14cm]{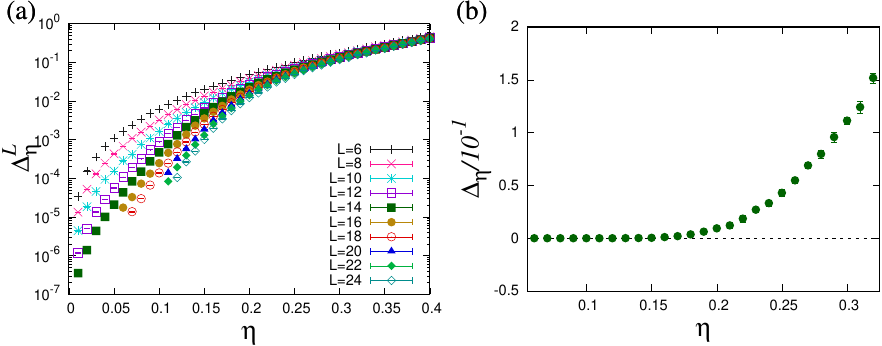}
\caption{Spectral gaps in the Floquet model. 
(a) Numerically obtained $\Delta_\eta^L$ as functions of the measurement strength $\eta$ for various $L$. 
(b) The spectral gap $\Delta_\eta$ in the thermodynamic limit obtained through the extrapolation of the numerical data. }
\label{fig:gap_Floquet-model}
\end{center}
\end{figure}

Figure \ref{fig:gap_Floquet-model} (a) shows the spectral gap $\Delta_\eta^L$ for various system sizes $L$ as functions of the measurement strength $\eta$. 
We can see that there is a transition between the gapped phase where $\Delta_\eta^L$ becomes almost independent of $L$ and the gapless phase where $\Delta_\eta^L$ decreases as $L$ is increased. 
This is more quantitatively illustrated in Fig. \ref{fig:gap_Floquet-model} (b); the spectral gap in the thermodynamic limit becomes $\Delta_\eta\simeq0$ when $\eta\lesssim0.18$ and it grows when $0.2\lesssim\eta$.

\begin{figure}[tb]
\begin{center}
    \includegraphics[width=15cm]{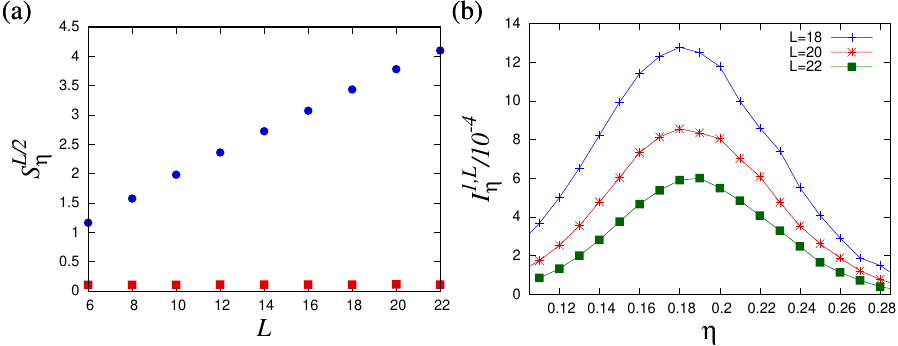}
\caption{The entanglement entropy and mutual information for the Floquet model. 
(a) The half-chain entanglement entropy $S^{L/2}_\eta$ as functions of $L$ with $\eta=0.12$ (blue circles) and $\eta=0.37$ (red squares). 
The time average is taken over $10^4$ steps after $t$ becomes larger than $\left|\ln(10^{-3})/\Delta_\eta^L\right|$. 
(b) The mutual information as functions of $\eta$, where the time average is taken over $10^5$ steps after $\left|\ln(10^{-3})/\Delta_\eta^L\right|$.  }
\label{fig:entanglement_Floquet-model}
\end{center}
\end{figure}
The spectral transition also corresponds to the ground-state entanglement transition of the effective Hamiltonian in the Floquet model. 
Figure \ref{fig:entanglement_Floquet-model} (a) shows the half-chain entanglement entropy averaged over time, $S^{L/2}_\eta$. 
The entanglement entropy becomes almost independent of the system size, $S^{L/2}_\eta\propto L^0$, in the gapped phase with large $\eta$. 
On the other hand, the entanglement entropy becomes proportional to the system size, $S^{L/2}_\eta\propto L$, in the gapless phase with small $\eta$. 
Figure \ref{fig:entanglement_Floquet-model} (b) shows a peak of the mutual information averaged over time, $I^{1,L}_\eta$, in the range $0.18\leq\eta_\mathrm{c}\leq0.2$, which corresponds to the threshold of the entanglement transition between the area-law and volume-law phases. 
Thus, the position of the peak is consistent with the spectral transition between the gapped and gapless phases as described in Fig. \ref{fig:gap_Floquet-model} (b).

\section{Discussions about the typical convergence}
\label{sec:typical-convergence}
As we have seen, the measurement-induced spectral transitions discussed in Sec. \ref{sec:numerics} rely on the typical convergence of the Lyapunov spectrum numerically found in Figs. \ref{fig:spectrum_temporally-random} and \ref{fig:spectrum_Floquet-model}. 
Therefore, to better understand the measurement-induced spectral transitions, it is important to discuss when the Lyapunov exponents exhibit the typical convergence in Eq. (\ref{eq:Lyapunov-spectrum_long-time_eta}).

To discuss why our models satisfy Eq. (\ref{eq:Lyapunov-spectrum_long-time_eta}), we consider the CPTP dynamics of density matrices averaged over outcomes,
\begin{align}
    \rho_{\tau+1}=\Gamma(\rho_\tau)
    =\sum_{\Omega}\mathcal{G}(\Omega) \rho_\tau \mathcal{G}^\dagger(\Omega),
    \label{eq:dynamics_averaged}
\end{align}
where $\sum_{\Omega}\mathcal{G}^\dagger(\Omega)\mathcal{G}(\Omega)=I$ is satisfied with $I$ being the identity matrix. 
In the case of our models, we have 
\begin{align}
\mathcal{G}(\Omega)=M_\eta(\{\zeta_{2,\ell }\})U(\{\theta_2^{\mu\nu}\})M_\eta(\{\zeta_{1,\ell }\})U(\{\theta_1^{\mu\nu}\})
\end{align}
with $\Omega=(\{\zeta_{1,\ell }\},\{\zeta_{2,\ell }\},\{\theta_1^{\mu\nu}\},\{\theta_2^{\mu\nu}\})$, where the summation represents 
\begin{align}
    \sum_\Omega=\sum_{\{\zeta_{1,\ell }\}}\sum_{\{\zeta_{2,\ell }\}}
    \prod_{t=1,2}\prod_{\mu,\nu=0}^3
    \int d\theta_t^{\mu\nu}g(\theta_t^{\mu\nu}).
\end{align}
Here, the probability densities of $\theta^{\mu\nu}$ are $g(\theta^{\mu\nu})=\Pi(\theta^{\mu\nu}/2\pi)/2\pi$ in the temporally random model and are the delta functions whose peaks are at values in Eqs. (\ref{eq:theta_1})-(\ref{eq:theta_3}) in the Floquet model, where $\Pi(\theta^{\mu\nu}/2\pi)$ is the rectangular function whose support is $-\pi\leq\theta^{\mu\nu}\leq+\pi$. 

Benoist et al.~\cite{benoist2019invariant} proved that if the stationary state $\rho_\infty=\Gamma(\rho_\infty)$ of the CPTP dynamics is unique and positive definite, the Lyapunov spectrum exhibits the typical convergence in Eq. (\ref{eq:Lyapunov-spectrum_long-time_eta}).
Thus, the unique and positive-definite stationary state of the CPTP dynamics ensures that we can analyze measurement-induced transitions through the well-defined Lyapunov spectrum. 
We note that measurement-induced transitions are observed in quantum trajectories but are usually invisible in the averaged CPTP dynamics, as is well known~\cite{li2018quantum,skinner2019measurement}. 
However, the discussions here indicate that properties of the CPTP dynamics play a crucial role for the well-definedness of the measurement-induced spectral transitions.

In the following, we discuss when stationary states satisfy these conditions from the properties of $\mathcal{G}(\Omega)$, while we skip the reason why $\rho_\infty$ satisfying these conditions leads to the typical convergence of the Lyapunov spectrum, which is detailed in Ref. \cite{benoist2019invariant}. 
We first explain the general condition of $\mathcal{G}(\Omega)$ for a CPTP dynamics to have a unique and positive-semidefinite stationary state in Sec. \ref{subsec:condition_stationary-state_general}. 
We then discuss it in our specific many-body models in Sec. \ref{subsec:condition_stationary-state_specific}, which is directly related to our numerical findings in Figs. \ref{fig:spectrum_temporally-random} and \ref{fig:spectrum_Floquet-model}.

\subsection{The condition for the unique and positive-definite stationary state}
\label{subsec:condition_stationary-state_general}
In this subsection, we give general discussions about the stationary states in the averaged dynamics and the typical convergence of the Lyapunov spectrum, not restricted to our models. 
In the dynamics described by Eq. (\ref{eq:dynamics_averaged}), there is always a positive semi-definite stationary state $\rho_\infty$, which satisfies
\begin{align}
    \Gamma(\rho_\infty)=\rho_\infty.
    \label{eq:stationary-state}
\end{align}
We discuss in what situations $\rho_\infty$ becomes unique and positive definite, which is the sufficient condition for the typical convergence of the Lyapunov exponents.

Conditions for the presence of a unique and positive-definite stationary state have been explored in discrete and continuous CPTP dynamics \cite{evans1977irreducible,frigerio1977quantum,frigerio1978stationary,burgarth2013ergodic,yoshida2024uniqueness}.
Here, we discuss a sufficient condition based on the discrete-time version of Ref. \cite{yoshida2024uniqueness}: if any operator $O$ can be constructed from $\{\mathcal{G}(\Omega)\}$ under multiplication, addition, and scalar multiplication, there is a unique and positive-definite stationary state. 
This condition is rewritten with using $\mathcal{V}({\bm \Omega}_\tau)=\mathcal{G}(\Omega_\tau)\cdots\mathcal{G}(\Omega_1)$ as
\begin{align}
    O=\sum_{\tau,{\bm \Omega}_\tau}z({\bm \Omega}_\tau)
    \mathcal{V}({\bm \Omega}_\tau),\ \forall\,O
    \label{eq:operator-construction}
\end{align}
where $z({\bm \Omega}_\tau)$ is a complex number that depends on the sequence of outcomes ${\bm \Omega}_\tau$. 
We give a proof of this statement based on Ref. \cite{yoshida2024uniqueness}. 

First, we give a proof of the positive definiteness using the assumption in Eq. (\ref{eq:operator-construction}). 
If an eigenvalue of $\rho_\infty$ were zero,
\begin{align}
    \bra{0}\rho_\infty\ket{0}=\bra{0}\Gamma^\tau(\rho_\infty)\ket{0}
    =\sum_{{\bm \Omega}_\tau}\left|\sqrt{\rho_\infty}
    \mathcal{V}^\dagger({\bm \Omega}_\tau)\ket{0}\right|^2=0
    \label{eq:zero-eigenvalue}
\end{align}
would be satisfied, where $\ket{0}$ is the corresponding eigenvector. 
Equation (\ref{eq:zero-eigenvalue}) leads to
\begin{align}
    \rho_\infty \mathcal{V}^\dagger({\bm \Omega}_\tau)\ket{0}=0,
    \label{eq:rho-V_zero}
\end{align}
for arbitrary $\tau$ and all possible ${\bm \Omega}_\tau$. 
Since any operator $O$ can be constructed from $\mathcal{V}^\dagger({\bm \Omega}_\tau)$, Eq. (\ref{eq:rho-V_zero}) indicates
\begin{align}
    \rho_\infty\ket{\psi}=0
    \label{eq:rho-psi_zero}
\end{align}
for any state $\ket{\psi}$, which would mean $\rho_\infty=0$. 
Thus, when Eq. (\ref{eq:operator-construction}) is satisfied, $\rho_\infty$ is always positive definite. 

Second, we give a proof of uniqueness. 
Suppose that there are two positive definite stationary states $\rho_\infty^1$ and $\rho_\infty^2$. 
Then, we consider the stationary state 
\begin{align}
    \rho_\infty(u)=(1-u)\rho_\infty^1-u\rho_\infty^2
    \label{eq:stationary-state_combination}    
\end{align}
with $0 \leq u \leq 1$. 
Since eigenvalues of $\rho_\infty^1$ and $\rho_\infty^2$ are positive, all eigenvalues of $\rho_\infty(u)$ are positive and negative with $u=0$ and $u=1$, respectively. 
Therefore, $\rho_\infty(u)$ becomes a positive semidefinite matrix with zero eigenvalue at a $u=u_0$ between $0$ and $1$, since the minimum eigenvalue of $\rho_\infty(u)$ should be zero in the range $0 \leq u \leq 1$. 
Thus, the above discussion about positive definiteness of the stationary state leads to $\rho_\infty(u_0)=0$ and thus $\rho_\infty^1\propto\rho_\infty^2$. 
This indicates that the stationary state is unique. 

\subsection{Discussions about our models}
\label{subsec:condition_stationary-state_specific}
We can theoretically show that the temporary random model explored in Sec. \ref{sec:numerics} exhibits the typical convergence of the Lyapunov exponents: the condition in Eq. (\ref{eq:operator-construction}) is satisfied and thus there is a unique and positive-definite stationary state. 
To this end, we prove that an arbitrary Pauli string can be obtained through multiplication, addition, and scalar multiplication of $\{\mathcal{G}(\Omega)\}$. 
This means that we can construct any operator from $\{\mathcal{G}(\Omega)\}$ since Pauli strings $\prod_{\ell =1}^L\sigma^{\mu_\ell }_\ell $ span the orthogonal basis in the space of $N \times N$ matrices with $N=2^L$. 
First, we find that $\sigma^0_\ell $ and $\sigma^3_\ell $ at an arbitrary position $\ell $ are obtained from the Kraus operators in Eq. (\ref{eq:Kraus-operator_each}), since they satisfy
\begin{align}
    \mathsf{M}_\eta(\zeta_\ell =+)
    +\mathsf{M}_\eta(\zeta_\ell =-)
    \propto\sigma^0_\ell ,\ \ 
    \mathsf{M}_\eta(\zeta_\ell =+)
    -\mathsf{M}_\eta(\zeta_\ell =-)
    \propto\sigma^3_\ell .
\end{align}
Second, we can also obtain $\sigma^1_\ell $ and $\sigma^2_\ell $ from $U(\{\theta_t^{\mu\nu}\})$ and $\sigma^3_\ell $ constructed above without affecting operators at other sites. 
This is because they can be made through 
\begin{align}
    U_2^+\sigma^3_\ell U_2^-\propto\sigma^1_\ell ,\ \     
    U_1^+\sigma^3_\ell U_1^-\propto\sigma^2_\ell ,
\end{align}
where $U_\nu^\pm=\prod_\ell \exp\left(\pm i\pi\sigma_\ell ^\nu/4\right)$ with $\nu=1,2$. 
Here, $U(\{\theta_1^{\mu\nu}\})=U_1^-,\,U(\{\theta_2^{\mu\nu}\})=U_1^+$ and $U(\{\theta_1^{\mu\nu}\})=U_2^-,\,U(\{\theta_2^{\mu\nu}\})=U_2^+$ are included in the ensemble of temporally random unitary matrices in Eqs. (\ref{eq:unitary-operator_each}) and (\ref{eq:unitary-operator_total}) with Eq. (\ref{eq:box-distribution}). 
Therefore, arbitrary Pauli operators at $\ell $ can be constructed through the procedure explained above. 
Repeating this operation, we can construct any Pauli string. 
Note that when local unitary gates for two spins are spatiotemporally random Haar unitary gates, which was explored in Ref. ~\cite{mochizuki2025measurement}, we can also confirm that Eq. (\ref{eq:operator-construction}) is satisfied. 
This is because the ensemble of spatiotemporal random Haar unitaries includes all Pauli strings that form the basis set for $N \times N$ matrices. 

Meanwhile, it is not evident that our Floquet model with spatiotemporally homogeneous unitary gates satisfies Eq. (\ref{eq:operator-construction}). 
However, it is likely that Eq. (\ref{eq:operator-construction}) holds in the Floquet model, since if the condition is satisfied that will be consistent with the typical convergence of the Lyapunov exponents found in Fig. \ref{fig:spectrum_Floquet-model}. 
We conjecture that any operator can be constructed from $\{\mathcal{G}(\Omega)\}$ even in the Floquet model, on the basis of two reasons written below. 
One is that our Floquet model is generic, i.e., the local Hamiltonian in Eq. (\ref{eq:unitary-operator_each}) includes all possible Pauli strings for two spins except the identity, and the parameters in Eqs. (\ref{eq:theta_1})-(\ref{eq:theta_3}) have no specific structure. 
The other is that any strong symmetry and resulting conserved quantity are absent, i.e., there is no unitary operator that commutes with all $\{\mathcal{G}(\Omega)\}$. 
This is contrasted to symmetric models where symmetries such as SU(2) symmetry \cite{majidy2023critical} and U(1) symmetry \cite{agrawal2022entanglement,barratt2022transitions,oshima2023charge} prevent monitored systems to satisfy Eq. (\ref{eq:operator-construction}); if $\{\mathcal{G}(\Omega)\}$ have a symmetry, $\sum_{\tau,{\bm \Omega}_\tau}z({\bm \Omega}_\tau)\mathcal{V}({\bm \Omega}_\tau)$ also have the symmetry, and thus operators that do not respect the symmetry cannot be made from $\{\mathcal{G}(\Omega)\}$.

\section{Summary}
\label{sec:summary}
We have explored the Lyapunov spectrum in monitored quantum systems with temporally random and Floquet unitary gates, where both gates have no spatial randomness. 
We have found that the Lyapunov spectrum becomes independent of measurement outcomes in both systems, which indicates that spatial and temporal randomness of unitary gates are not essential for the typical convergence of the Lyapunov spectrum.  
We have analytically shown that the temporally random model exhibits a unique and positive-definite stationary state in the CPTP dynamics averaged over outcomes, which leads to the typical convergence of the Lyapunov spectrum. 
This highlights that properties of the averaged CPTP dynamics are crucial in defining measurement-induced spectral transitions through the Lyapunov spectrum of typical trajectories, although whether measurement-induced transitions actually occur or not is invisible from the CPTP dynamics~\cite{li2018quantum,skinner2019measurement}. 
In addition, we have explored transitions of the spectral gap and entanglement entropy. 
In both monitored systems, the spectral transition between the gapped and gapless phases corresponds to the entanglement transition of the ground states of the effective Hamiltonians. 
This coincidence is analogous to that in isolated systems, where vanishing spectral gaps indicate phase transitions of ground states. 
While the coincidence of the spectral and entanglement transitions was observed in monitored systems with spatiotemporally random unitary gates \cite{mochizuki2025measurement}, our models explored in this work are less random in that unitary gates are spatially homogeneous. 
Thus, our results suggest that the spectrum and the ground-state entanglement are related in various systems with spatial uniformity. 
For example, it should be an intriguing future work to explore the relation between the spectrum and entanglement in monitored dynamics where unitary dynamics are generated by Hamiltonians extensively studied in condensed matter physics, like the XXZ Hamiltonian and the Hubbard Hamiltonian \cite{fuji2020measurement,tang2020measurement}. 

\section*{Data availability statement}
The data that support the findings of this study are openly available at the following URL/DOI:
https://zenodo.org/records/15532381.

\section*{acknowledgements}
We thank Yohei Fuji and Hisanori Oshima for fruitful discussions. 
We also thank Hironobu Yoshida for discussions in Sec. \ref{sec:typical-convergence}. 
This work was supported by JST ERATO Grant Number JPMJER2302, Japan. 
K.M. was supported by JSPS KAKENHI Grant Number JP23K13037. 
R.H. was supported by JSPS KAKENHI Grant Number JP24K16982.

\appendix
\section{Justification of Eq. (\ref{eq:spectrum-candidates_limit})}
\label{sec:justification_spectrum}
We give an explanation why Eq. (\ref{eq:spectrum-candidates_limit}) is satisfied \cite{ershov1998concept}. 
To this end, we consider the $N \times q$ matrix $Y_q({\bm \omega}_t)$ defined as
\begin{align}
    Y_q({\bm \omega}_t)
    &=\left[\ket{\phi_1({\bm \omega}_t)},\ket{\phi_2({\bm \omega}_t)},\cdots,\ket{\phi_q({\bm \omega}_t)}\right]\nonumber\\
    &=\left[V({\bm \omega}_t)\ket{\tilde{\Psi}_1^0},
    V({\bm \omega}_t)\ket{\tilde{\Psi}_2^0},\cdots,
    V({\bm \omega}_t)\ket{\tilde{\Psi}_q^0}\right],
    \label{eq:state-matrix_bare}
\end{align}
where $t=sb$ and $V({\bm \omega}_t)=V(\tilde{\bm \omega}_s)V(\tilde{\bm \omega}_{s-1}) \cdots V(\tilde{\bm \omega}_1)$. 
The procedure in Eqs. (\ref{eq:evolution_tilde})-(\ref{eq:normalization}) corresponds to the QR decomposition of $Y_q({\bm \omega}_t)$,
\begin{align}
    Y_q({\bm \omega}_t)=Q({\bm \omega}_t)R({\bm \omega}_t),
    \label{eq:QR-decomposition}
\end{align}
where $Q({\bm \omega}_t)$ is the $N \times q$ matrix composed of $\{\ket{\tilde{\Psi}_i({\bm \omega}_t)}\}$,
\begin{align}
    Q({\bm \omega}_t)
    =\left[\ket{\tilde{\Psi}_1({\bm \omega}_t)},\ket{\tilde{\Psi}_2({\bm \omega}_t)},\cdots,\ket{\tilde{\Psi}_q({\bm \omega}_t)}\right],
    \label{eq:state-matrix_orthonormalized}
\end{align}
and $R({\bm \omega}_t)$ is the $q \times q$ upper triangular matrix whose elements are
\begin{align}
    R_{ij}({\bm \omega}_t)
    =\left\{\begin{array}{cc}
        \langle\tilde{\Psi}_i({\bm \omega}_t)|
        \phi_j({\bm \omega}_t)\rangle&(i=1,2,\cdots,j-1)\\
        \prod_{r=1}^s\sqrt{\langle\chi_i({\bm \omega}_{rb})|
        \chi_i({\bm \omega}_{rb})\rangle}&(i=j)\\
        0&(i=j+1,\cdots,q)
    \end{array}\right.
    \label{eq:triangle_matrix}
\end{align}
This is obtained by applying the QR decomposition to $V(\tilde{\bm \omega}_s)Q({\bm \omega}_{(s-1)b})$,  which leads to $V(\tilde{\bm \omega}_s)Q({\bm \omega}_{(s-1)b})=Q({\bm \omega}_{sb})\tilde{R}({\bm \omega}_{sb})$.
Here, $\tilde{R}({\bm \omega}_{sb})$ is the upper triangular matrix whose diagonal elements are $\sqrt{\langle\chi_i({\bm \omega}_{sb})|\chi_i({\bm \omega}_{sb})\rangle}$. 
Thus, the diagonal elements of $R({\bm \omega}_{sb})$ become $R_{ii}({\bm \omega}_{sb})=\tilde{R}_{ii}({\bm \omega}_{sb})\tilde{R}_{ii}({\bm \omega}_{(s-1)b}) \cdots \tilde{R}_{ii}({\bm \omega}_b)$. 
From Eq. (\ref{eq:triangle_matrix}), we can understand that the candidates of the Lyapunov exponents in Eq. (\ref{eq:Lyapunov-spectrum_each_separate}) can be written as
\begin{align}
    \tilde{\varepsilon}_i({\bm \omega}_t)
    =-\frac{1}{t}\ln\left[R_{ii}({\bm \omega}_t)\right].
    \label{eq:Lyapunov-spectrum_each_once}
\end{align}
Therefore, we will confirm Eq. (\ref{eq:spectrum-candidates_limit}) through evaluating $Y_q({\bm \omega}_t)$ in the long-time regime.  

We expand the initial states $\{\ket{\tilde{\Psi}_i^0}\}$ through the orthonormal basis $\{\ket{\Phi_j({\bm \omega}_t)}\}$,
\begin{align}
    \ket{\tilde{\Psi}_i^0}
    =\sum_{j=1}^Nc_{ij}({\bm \omega}_t)\ket{\Phi_j({\bm \omega}_t)}.
    \label{eq:expansion}
\end{align}
Applying $V({\bm \omega}_t)$ to $\ket{\tilde{\Psi}_i^0}$, we obtain
\begin{align}
    \ket{\phi_i({\bm \omega}_t)}
    =V({\bm \omega}_t)\ket{\tilde{\Psi}_i^0}
    =\sum_{j=1}^Nc_{ij}({\bm \omega}_t)\Lambda_j({\bm \omega}_t)
    \ket{\Psi_j({\bm \omega}_t)}.
    \label{eq:expansion_phi_1}
\end{align}
Then, we evaluate $Z_q({\bm \omega}_t)$ which is defined as
\begin{align}
    Z_q({\bm \omega}_t)
    &=\det\left[Y_q^\dagger({\bm \omega}_t)
    Y_q({\bm \omega}_t)\right]
    =\sum_\Xi\mathrm{sign}(\Xi)\prod_{i=1}^q\left[
    \sum_{j=1}^Nc_{ij}^*({\bm \omega}_t)\Lambda_j^2({\bm \omega}_t)
    c_{\Xi(i)j}({\bm \omega}_t)\right],
    \label{eq:Gamma_definition}
\end{align}
where $\Xi$ represents a permutation of $(1,2,\cdots,q)$. 
In the right-hand side of Eq. (\ref{eq:Gamma_definition}), terms including $\Lambda_j^2({\bm \omega}_t)\Lambda_j^2({\bm \omega}_t)$ do not appear since they cancel out due to $\mathrm{sign}(\Xi)$. 
All the other realizations $\Lambda_{\Upsilon(1)}^2({\bm \omega}_t)\Lambda_{\Upsilon(2)}^2({\bm \omega}_t)\cdots\Lambda_{\Upsilon(q)}^2$ emerge in the sum, where $\Upsilon=[\Upsilon(1),\Upsilon(2),\cdots,\Upsilon(q)]$ is a set of integers which satisfies $1 \leq \Upsilon(1) < \Upsilon(2) \cdots < \Upsilon(q-1) < \Upsilon(q) \leq N$. 
Thus, $Z_q({\bm \omega}_t)$ can be written as
\begin{align}
    Z_q({\bm \omega}_t)
    &=\sum_\Upsilon\Lambda_{\Upsilon(1)}^2({\bm \omega}_t)
    \Lambda_{\Upsilon(2)}^2({\bm \omega}_t)
    \cdots\Lambda_{\Upsilon(q)}^2({\bm \omega}_t)
    \sum_\Xi\mathrm{sign}(\Xi)
    \prod_{i=1}^q\sum_{j=1}^q
    c_{i\Upsilon(j)}^*({\bm \omega}_t)
    c_{\Xi(i)\Upsilon(j)}({\bm \omega}_t)
    \nonumber\\&=\sum_\Upsilon
    \left|\Lambda_{\Upsilon(1)}({\bm \omega}_t)
    \Lambda_{\Upsilon(2)}({\bm \omega}_t)
    \cdots\Lambda_{\Upsilon(q)}({\bm \omega}_t)
    D_\Upsilon({\bm \omega}_t)\right|^2,
    \label{eq:Gamma_determinant}
\end{align}
where
\begin{align}
    D_\Upsilon({\bm \omega}_t)
    =\det\left(\left[\begin{array}{cccc}
    c_{1\Upsilon(1)}({\bm \omega}_t) &
    c_{1\Upsilon(2)}({\bm \omega}_t) &
    \cdots &
    c_{1\Upsilon(q)}({\bm \omega}_t)\\
    c_{2\Upsilon(1)}({\bm \omega}_t) &
    c_{2\Upsilon(2)}({\bm \omega}_t) &
    \cdots&\cdots\\
    \cdots&\cdots&\cdots&\cdots\\
    c_{q\Upsilon(1)}({\bm \omega}_t) &
    \cdots & \cdots &
    c_{q\Upsilon(q)}({\bm \omega}_t)
    \end{array}\right]\right).
    \label{eq:determinant}
\end{align}
From Eq. (\ref{eq:Gamma_determinant}), we can obtain the lower and upper bounds of $Z_q({\bm \omega}_t)/\left[\Lambda_1({\bm \omega}_t)\Lambda_2({\bm \omega}_t)\cdots\Lambda_q({\bm \omega}_t)\right]^2$,
\begin{align}
    \left|D_{12 \cdots q}({\bm \omega}_t)\right|^2
    \leq\frac{Z_q({\bm \omega}_t)}
    {\left[\Lambda_1({\bm \omega}_t)\Lambda_2({\bm \omega}_t)
    \cdots\Lambda_q({\bm \omega}_t)\right]^2}
    \leq\sum_\Upsilon\left|D_\Upsilon({\bm \omega}_t)\right|^2=1.
    \label{eq:inequality_Gamma}
\end{align}
Since $Z_q({\bm \omega_t})$ can be written as $Z_q({\bm \omega_t})=\det\left[R^\dagger({\bm \omega}_t)R({\bm \omega}_t)\right]=\prod_{i=1}^qR_{ii}^2({\bm \omega}_t)$, $R_{qq}({\bm \omega}_t)$ becomes $R_{qq}({\bm \omega}_t)=\sqrt{Z_q({\bm \omega}_t)/Z_{q-1}({\bm \omega}_t)}$. 
Thus, Eq. (\ref{eq:inequality_Gamma}) leads to
\begin{align}
    \left|D_{12 \cdots q}({\bm \omega}_t)\right|\Lambda_q({\bm \omega}_t)
    \leq R_{qq}({\bm \omega}_t) \leq
    \frac{\Lambda_q({\bm \omega}_t)}{\left|D_{12 \cdots q-1}({\bm \omega}_t)\right|}.
    \label{eq:inequality_R}
\end{align}
Both $\{\ket{\tilde{\Psi}_i^0}\}$ and $\{\ket{\Phi_j({\bm \omega}_t)}\}$ span  orthonormal basis sets, which means $\left|D_{12 \cdots q}({\bm \omega}_t)\right|=O(t^0)$ since $\{\ket{\tilde{\Psi}_i^0}\}$ are randomly chosen initial states independent of ${\bm \omega}_t$. 
Therefore, Eq. (\ref{eq:inequality_R}) indicates that $\tilde{\varepsilon}_q({\bm \omega}_t)$ in Eq. (\ref{eq:Lyapunov-spectrum_each_once}) approaches $-\ln[\Lambda_q({\bm \omega}_t)]/t=\varepsilon_q({\bm \omega}_t)$ for sufficiently large $t$. 
Since $q$ can be an arbitrary integer in the range $1 \leq q \leq N$, Eq. (\ref{eq:spectrum-candidates_limit}) is satisfied for all $i=1,2,\cdots,N$. 

\section{Justification of Eq. (\ref{eq:state-candidates_limit})}
\label{sec:justification_state}
We show Eq. (\ref{eq:state-candidates_limit}), that is, $\ket{\tilde{\Psi}_i({\bm \omega}_t)}$ approaches $\ket{\Psi_i({\bm \omega}_t)}$ for large $t$. 
To this end, we expand $\ket{\tilde{\Psi}_i({\bm \omega}_t)}$ as
\begin{align}
    \ket{\tilde{\Psi}_i({\bm \omega}_t)}
    =\sum_{j=1}^NC_{ij}({\bm \omega}_t)
    \ket{\Psi_j({\bm \omega}_t)},
    \label{eq:expansion_psi}
\end{align}
with $1 \leq i \leq N$. 
From Eqs. (\ref{eq:state-matrix_bare})-(\ref{eq:state-matrix_orthonormalized}) and (\ref{eq:expansion_psi}), we can understand that $\ket{\phi_i({\bm \omega}_t)}$ becomes 
\begin{align}
   \ket{\phi_i({\bm \omega}_t)}
   =\sum_{k=1}^iR_{ki}({\bm \omega}_t)
   \ket{\tilde{\Psi}_k({\bm \omega}_t)}
   =\sum_{j=1}^N\left[\sum_{k=1}^iC_{kj}({\bm \omega}_t)
   R_{ki}({\bm \omega}_t)\right]\ket{\Psi_j({\bm \omega}_t)}.
   \label{eq:expansion_phi_2}
\end{align}
Comparing Eqs. (\ref{eq:expansion_phi_1}) and (\ref{eq:expansion_phi_2}), we can obtain
\begin{align}
    C_{ij}({\bm \omega}_t)
    =\frac{\Lambda_j({\bm \omega}_t)
    c_{ij}({\bm \omega}_t)-
    \sum_{k=1}^{i-1}C_{kj}({\bm \omega}_t)
    \sum_{n=1}^N\Lambda_n({\bm \omega}_t)
    c_{in}({\bm \omega}_t)
    C_{kn}^*({\bm \omega}_t)}
    {R_{ii}({\bm \omega}_t)},
    \label{eq:coefficient}
\end{align}
where $R_{ki}({\bm \omega}_t)=\langle\tilde{\Psi}_k({\bm \omega}_t)|\phi_i({\bm \omega}_t)\rangle=\sum_{n=1}^NC_{kn}^*({\bm \omega}_t)c_{in}({\bm \omega}_t)\Lambda_n({\bm \omega}_t)$ is used. 

Now, for large $t$, we show 
\begin{align}
    |C_{ij}({\bm \omega}_t)|\lesssim
    \begin{cases}
        \Lambda_j({\bm \omega}_t)/\Lambda_i({\bm \omega}_t)&(i \leq j)\\
        \Lambda_i({\bm \omega}_t)/\Lambda_j({\bm \omega}_t)&(i \geq j)
    \end{cases}
    \label{eq:hypothesis}
\end{align}
using the inductive method. 
In the following, we approximate the diagonal elements $R_{ii}({\bm \omega}_t)$ and the singular values $\Lambda_i({\bm \omega}_t)$ as $R_{ii}({\bm \omega}_t)\simeq\Lambda_i({\bm \omega}_t)\simeq\exp(-\varepsilon_it)$, on the basis of Eqs. (\ref{eq:inequality_R}) and (\ref{eq:Lyapunov-spectrum_long-time}). 
If Eq. (\ref{eq:hypothesis}) is satisfied, $C({\bm \omega}_t)$ becomes a unitary and diagonal matrix asymptotically and thus $\ket{\tilde{\Psi}_i({\bm \omega}_t)}$ approaches $\ket{\Psi_i({\bm \omega}_t)}$. This is because $\Lambda_j({\bm \omega}_t)/\Lambda_i({\bm \omega}_t)\simeq\exp\left[-(\varepsilon_j-\varepsilon_i)t\right]$ with $\varepsilon_i<\varepsilon_j$ exhibits exponential decay when $i < j$. 
If we focus on $i=1$, we can easily confirm that Eq. (\ref{eq:hypothesis}) is satisfied, since the second term in the right-hand side of Eq. (\ref{eq:coefficient}) is absent and thus $C_{1j}({\bm \omega}_t)$ becomes
\begin{align}
    C_{1j}({\bm \omega}_t)\simeq
    \frac{\Lambda_j({\bm \omega}_t)}{\Lambda_1({\bm \omega}_t)}
    c_{1j}({\bm \omega}_t)
    \simeq\exp[-(\varepsilon_j-\varepsilon_1)t]c_{1j}({\bm \omega}_t).
    \label{eq:hypothesis_i=1}
\end{align}
Then, we evaluate $C_{mj}({\bm \omega}_t)$ with $m\geq2$ under the assumption that Eq. (\ref{eq:hypothesis}) is satisfied for $i=1,2,\cdots,m-1$. 
In this case, $C({\bm \omega}_t)$ becomes a block diagonal matrix for large $t$, where two sections $1 \leq i,j \leq m-1$ and $m \leq i,j \leq N$ are separated, owing to the unitarity of $C({\bm \omega}_t)$. 
When $j > m$, $C_{mj}({\bm \omega}_t)$ can be evaluated as
\begin{align}
    C_{mj}({\bm \omega}_t)&\simeq e^{-(\varepsilon_j-\varepsilon_m)t}
    c_{mj}({\bm \omega}_t)
    -\sum_{k=1}^{m-1}
    \left[\sum_{n=1}^ke^{-(\varepsilon_j-\varepsilon_m)t}
    c_{mn}({\bm \omega}_t)
    +\sum_{n=k+1}^Ne^{-(2\varepsilon_n-2\varepsilon_k+\varepsilon_j-\varepsilon_m)t}
    c_{mn}({\bm \omega}_t)\right]\nonumber\\
    &\simeq0,
    \label{eq:evaluation_I<j}
\end{align}
since all terms in the right-hand side exhibit exponential decay with respect to $t$. Equation (\ref{eq:evaluation_I<j}) indicates that $C_{mm}({\bm \omega}_t)$ becomes an eigenvalue of $C({\bm \omega}_t)$ and thus $C_{jm}({\bm \omega}_t)$ with $j>m$ should also be zero owing to the unitarity $C^\dagger({\bm \omega}_t)C({\bm \omega}_t)=I$. 
Therefore, in the long-time regime, Eq. (\ref{eq:hypothesis}) is satisfied for $i=1,2,\cdots,m$ with arbitrary $m$, which means that Eq. (\ref{eq:state-candidates_limit}) is satisfied for arbitrary $i=1,2,\cdots,N$.  

\section*{Reference}
\bibliographystyle{iopart-num}
\bibliography{reference.bib}

\end{document}